\documentclass[twocolumn,showpacs,amsmath,pra,aps,amssymb,superscriptaddress]{revtex4-1} 
\usepackage[T1]{fontenc}
\usepackage[latin9]{inputenc}
\usepackage{times}
\usepackage{color} 
\usepackage{xspace}
\usepackage{amssymb,amsmath}
\usepackage{amsbsy}
\usepackage[pdftex]{graphicx}
\usepackage{bm}
\usepackage{float}

\usepackage[unicode,breaklinks]{hyperref}
\hypersetup{
    unicode=true,
    plainpages=false, 
    colorlinks=true,
    linkcolor=blue,
    citecolor=blue,
    filecolor=black,
    urlcolor=blue
}
\usepackage{url}
\usepackage{verbatim}

\newcommand{\br}{\mathbf{r}}

\begin{document}

\title{Properties of a dipolar condensate with three-body interactions} 
\author{P.~B.~Blakie}  
\affiliation{Department of Physics, Centre for Quantum Science, and Dodd-Walls Centre for Photonic and Quantum Technologies, University of Otago, Dunedin, New Zealand}

\begin{abstract}
We obtain the phase diagram for a harmonically trapped dilute dipolar condensate with a short ranged conservative three-body interaction. We show that this system supports two distinct fluid states: a usual condensate state and a self-cohering droplet state. We develop a simple model to quantify the energetics of these states, which we verify with full numerical calculations. Based on our simple model we develop a phase diagram showing that there is a first order phase transition between the states. Using dynamical simulations we explore the phase transition dynamics, revealing that the droplet crystal observed in previous work is an excited state that arises from heating as the system crosses the phase transition. Utilising our phase diagram we show it is feasible to produce a single droplet by dynamically adjusting the confining potential.
\end{abstract}
 

\maketitle

\section{Introduction}
Quantum gases with significant dipole moments are an interesting playground for exploring the role of long-ranged interactions on superfluidity and spontaneous crystallization in a quantum fluid \cite{Ronen2007a,Buchler2007a,Lahaye2009a,Boninsegni2012a,Moroni2014a,Lu2015a}. In the regime of dominant dipole-dipole interactions (DDIs), where crystallization might be expected to occur, dilute gases are fragile to local mechanical collapse \cite{Komineas2007a,Koch2008a, WIlson2009a,Parker2009a,Lahaye2009a,Linscott2014a}. In order to stabilize this system an effective interaction is required that can balance the tendency of the dominant DDI to collapse the system towards infinite density spikes. A repulsive short-ranged three-body interaction (TBI) \cite{Kohler2002a,Braaten2002a,Bulgac2002a,Buchler2007b} meets these requirements: it produces an energy contribution that increases with $n^3$, where $n$ is the number density, and thus dominates over the two-body DDI as the density increases. A theoretical proposal has shown how to produce a repulsive TBI in a dilute gas of polar molecules \cite{Petrov2014a}. It is also expected that significant TBIs could emerge in the vicinity of Feshbach resonances used to modify the s-wave scattering length \cite{Kohler2002a,Braaten2002a}. Indeed, some evidence for such interactions has been presented in experiments with $^{85}$Rb \cite{Everitt2015a}. 

\begin{figure}[htbp]
   \centering
 \includegraphics[width=2.5in]{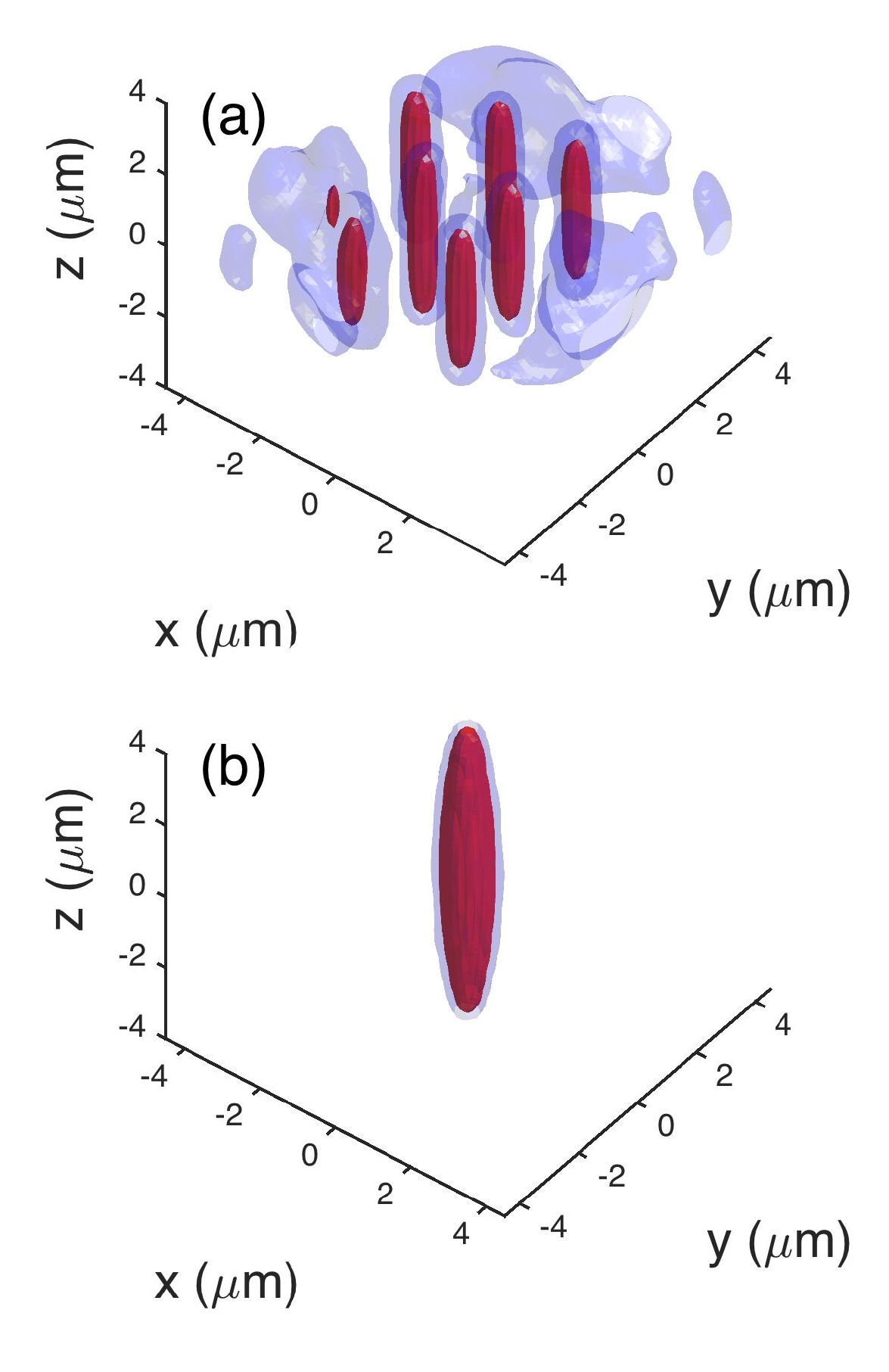} 
  \vspace*{-0.5cm}
   \caption{(Color
 online)  Examples of  (a) a crystal of droplets and (b) a single self-cohering droplet that can be produced when a dilute dipolar gas with three-body interactions is taken through different paths into the droplet phase.
 Red surface indicates a high-density isosurface at $n=2\times10^{20}\,$m$^{-3}$ and the light-blue surface indicates a low-density isosurface at $n=0.2\times10^{20}\,$m$^{-3}$. Parameters and these results are discussed further in Fig.~\ref{paths}. }
   \label{densurf}
\end{figure}

In this paper we consider the ground state properties of a dilute gas of dipolar atoms with an appreciable TBI. We show that this system has low-density and high-density ground states. The low density states are typical condensate states, with properties largely determined by the two-body interactions (DDI and the s-wave contact interaction) and the external confining potential. The high-density state, which can occur when the DDI dominates over the two-body contact interaction, is a self-cohering droplet in which the attractive DDI is balanced by the repulsive TBI. These  states are self-cohering in the sense that they are stable even when the confinement in the plane transverse to the orientation of the dipole moments is removed. Previously, such self-cohering droplets (or quasi-two-dimensional bright solitons) have been predicted for dipolar condensates with negatively turned dipoles \cite{Pedri2005a}. We show that the transition between the low- and high-density states occurs in oblately confined traps via a first order phase transition.  We also show that depending on how that transition is crossed, either a crystal of droplets or a single droplet can be produced, as shown in Fig.~\ref{densurf}.

This work is also motivated by a recent experiment with $^{164}$Dy that observed the formation of a droplet crystal. A full understanding of the key physics behind this observation has yet to be developed. Two groups have simulated the formation dynamics by augmenting the standard meanfield description of this system with a TBI \cite{Xi2016a,Bisset2015a}. 
Other recent work \cite{Ferrier-Barbut2016a,Wachtler2016a} has presented evidence that quantum fluctuations effects are important in stabilizing the droplets. The detailed microscopic understanding of both proposed mechanisms is unclear since: i) Dysprosium atoms have complex collisional properties (e.g.~see \cite{Maier2015a}) and there are no quantitative predictions for the magnitude of the expected TBI; ii) there is limited understanding of quantum fluctuation effects in microscopic and highly anisotropic droplets in the dipole dominated regime (noting that \cite{Ferrier-Barbut2016a} has extrapolated a result \cite{Schatzhold2006a,Lima2011a,Lima2012a} for homogeneous condensate with weak dipoles). Our results here provide quantitative predictions for the role of TBIs and thus may be useful in determining whether these interactions play a role in the aforementioned system.

\section{Formalism}
Here we introduce the basic formalism for the dynamics and equilibrium states of a dipolar condensate with TBIs. We focus here on the case of an atomic gas with magnetic dipoles, although we emphasize that these results will also apply to systems of polar molecules.
\subsection{Meanfield theory}
The evolution of the system is  described by the time-dependent Gross-Pitaevskii equation (GPE)
 \begin{align}
i\hbar\frac{\partial \psi}{\partial t}=& \mathcal{L}_{\mathrm{GP}}\psi,\\
=&\left[H_{\mathrm{sp}}\!+\!\int\!d\br'\,U(\br\!-\!\br')|\psi(\br')|^2+\frac{\kappa_3}{2}|\psi|^4\right]\psi,\label{tdGPE}
\end{align}
where
\begin{equation}
H_{\mathrm{sp}}=-\frac{\hbar^2\nabla^2}{2m}+V_{\mathrm{trap}}(\br),
\end{equation}
is the single particle Hamiltonian,  $m$ is the atomic mass and $V_{\mathrm{trap}}$ is the external trap potential. The condensate wavefunction $\psi$ is taken to be normalized to the number of particles $N$. The  two-body contact interaction and the DDI are described by the term
\begin{align}
U(\mathbf{r})=\frac{4\pi\hbar^2a_s}{m}\delta(\br)+\frac{\mu_0\mu^2}{4\pi}\frac{1-3\cos^2\theta}{r^3},
\end{align}
where the dipoles (of magnetic moment $\mu$) are taken to be polarized along $z$ and $\theta$ is the angle between $\br$ and the $z$-axis.  The two-body contact interaction, parameterized by the $s$-wave scattering length  $a_s$, which can be changed using a magnetic Feshbach resonance. The last term in (\ref{tdGPE}) describes the short-ranged TBIs. In general the coefficient $\kappa_3$ of this term is complex, with  the real part characterizing the strength of the conservative interaction and the imaginary part quantifying the three-body recombination loss rate.  Here we will only consider $\mathrm{Im}\{\kappa_3\}=0$ so that the ground states are indefinitely stable (do not decay through loss). In practice our predictions only require that the loss rate is sufficiently small compared to the inverse timescales for the relevant conservative dynamics (this is the case in the experiments reported in Ref.~\cite{Kadau2016a}).
  
The stationary states satisfy the  time-independent GPE
\begin{equation}
\mu_c\psi_0= \mathcal{L}_{\mathrm{GP}}\psi_0,\label{tiGPE}
\end{equation}
where $\mu_c$ is the chemical potential. We note that this equation can also be obtained as the condition for an extrema to the energy functional
\begin{equation}
E\!=\!\int\!dr\,\psi^*\!\left[H_{\mathrm{sp}}\!+\frac{1}{2}\!\int\!d\br'\,U(\br\!-\!\br')|\psi(\br')|^2+\frac{\kappa_3}{6}|\psi|^4\right]\!\psi,\label{Efunc}
\end{equation}
with $\psi$ constrained to be normalized to $N$ atoms and $\mu_c$ the associated Lagrange multiplier for this constraint,

Here we solve for stationary solutions to Eq.~(\ref{tiGPE}) in a cylindrically symmetric trap of the form
\begin{equation}
V_{\mathrm{trap}}(\br) =\frac{1}{2}m(\omega_\rho^2\rho^2+\omega_z^2z^2),
\end{equation}
 where $\rho=\sqrt{x^2+y^2}$ is the radial coordinate, and $\{\omega_\rho,\omega_z\}$ are the trap angular frequencies. In this case the ground states are also cylindrically symmetric and we can use an extension of the technique developed in Ref.~\cite{Ronen2006a} (and applied by us in Refs.~\cite{Bisset2012,Bisset2013b}) to account for the TBI. 

\subsection{Variational treatment}
Since  directly solving for ground states requires large scale numerical techniques it is also of interest to develop a simple variational approach.  Here we do this via a  Gaussian ansatz in which the condensate is described by the two variational width parameters $\{w_\rho,w_z\}$ as
\begin{equation}
\psi_{\mathrm{var}}(\br)=\sqrt{\frac{8N}{\pi^{3/2}w_{\rho}^{2}w_{z} }}\exp\left[-2\left(\frac{\rho^{2}}{w_{\rho}^{2}}+\frac{z^{2}}{w_{z}^{2}}\right)\right].
\end{equation}
Evaluating the energy functional (\ref{Efunc}) with this ansatz yields
  \begin{align}
  E&(w_\rho,w_z)= {{N\hbar\omega_z} \left(\frac{2a_z^2}{w_{\rho}^{2}}+\frac{a_z^2}{w_{z}^{2}}\right)}+ {\frac{N\hbar\omega_z}{16}\left(\frac{2}{\lambda^2}\frac{w_{\rho}^{2}}{a_z^2}+\frac{w_{z}^{2}}{a_z^2}\right)}\nonumber \\
  &+ {\frac{8N^{2}\hbar{\omega_z}a_z^3}{\sqrt{2\pi}w_{\rho}^{2}w_{z}}\left[\frac{a_s}{a_z}-\frac{a_{\mathrm{dd}}}{a_z}f\left(\frac{w_{\rho}}{w_{z}}\right)\right]} + {\frac{32N^{3}\kappa_3}{9\sqrt{3}\pi^{3}w_{\rho}^{4}w_{z}^{2}}} \label{Evar},
 \end{align} 
 where we have introduced the dipole length $a_{\mathrm{dd}}\equiv\mu_0\mu^2m/12\pi\hbar^2$ and the function  
 \begin{align}
 f(x)=\frac{1+2x^2}{1-x^2}-\frac{3x^2\mathrm{arctanh}\sqrt{1-x^2}}{(1-x^2)^{3/2}}.
 \end{align}
 
 \section{Results}
 \begin{figure}[htbp]
   \centering
 \includegraphics[width=3.35in]{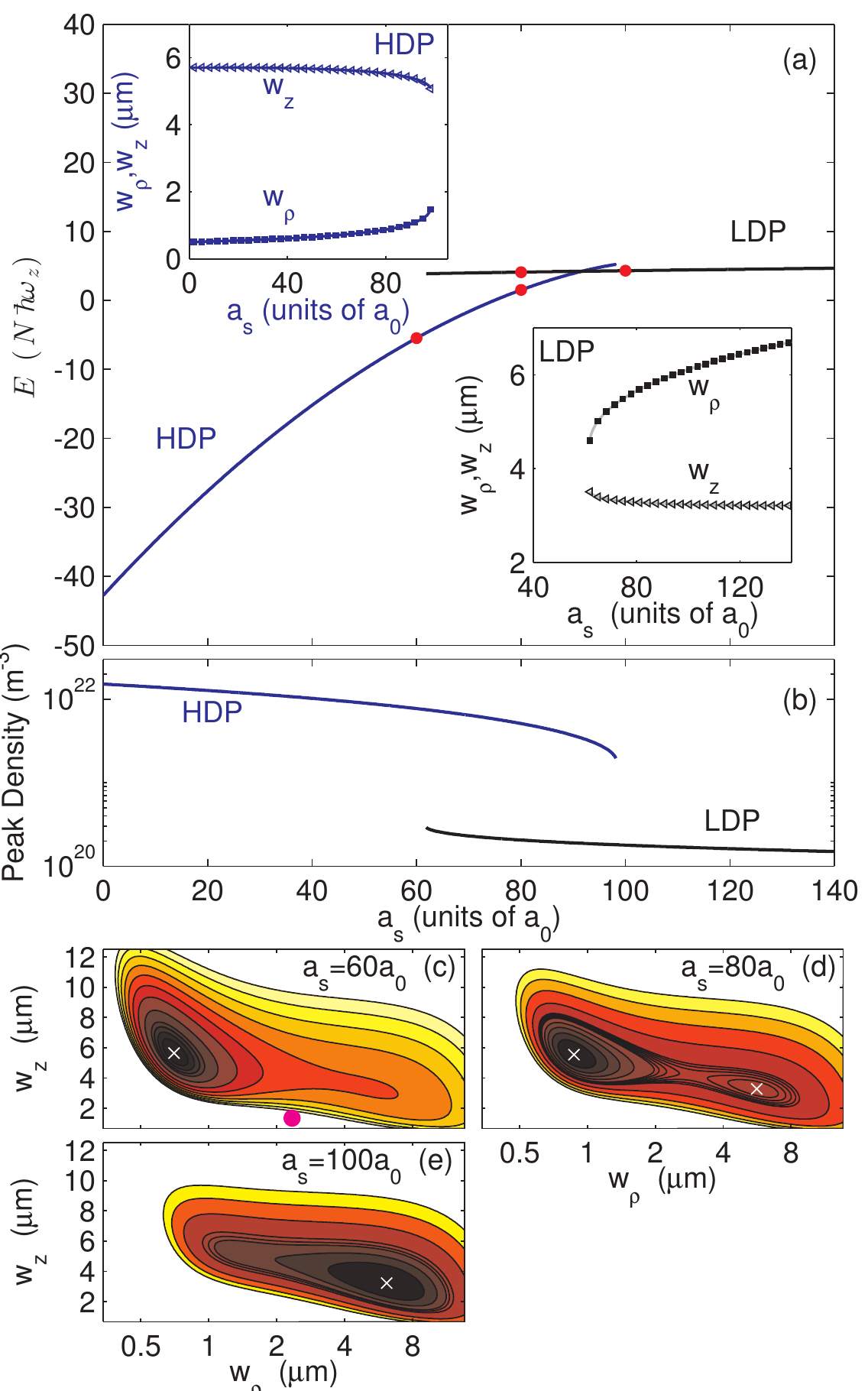} 
  \vspace*{-0.25cm}
   \caption{(Color
 online)  Variational solution properties. (a) Energies of the minima found in the variational solution as a function of s-wave scattering length. Up to two minima are found and these are represented by the blue line branch (labelled HDP) and the black line branch (labelled LDP). The red dots indicate the values of $a_s$ where the energy surface is evaluated in subplots (c)-(e). Insets show the values of the width parameters $\{w_\rho,w_z\}$ on the HDP branch (left inset) and the LDP branch (right inset). (b) The corresponding peak density  $n_{\mathrm{peak}}=|\psi_{\mathrm{var}}(\mathbf{0})|^2$ of the solutions on each branch. (c)-(e) shows energy contours of the variational solution energy as a function of $w_\rho$ and $w_z$. The local minima are indicated by small white crosses [corresponding to the energies indicated by the red dots in (a)]. In (c) the solution for the non-interacting case is shown as a pink circle for reference. Parameters are for $N=15\times10^3$ $^{164}$Dy atoms in a trap with $\{\omega_\rho,\omega_z\}=2\pi\times\{45,133\}\,$s$^{-1}$, and $\kappa_3=5.87\times10^{-39}\hbar$m$^6/$s (also see \cite{Bisset2015a}).}
   \label{Esurf}
\end{figure}

 \subsection{Ground and metastable states}
 The properties of the variational solution are explored in Fig.~\ref{Esurf} as the s-wave scattering length is varied.
For each value of $a_s$ we find either one or two local minima [see Figs.~\ref{Esurf}(c)-(e)], and the associated stable (or meta-stable) states are observed to lie on two distinct solution branches. These two branches have different energy character [Fig.~\ref{Esurf}(a)], but also differ physically is that one branch is highly prolate [i.e.~$w_z\gg w_\rho$, see left inset to Fig.~\ref{Esurf}(a)] and of relatively high density [Fig.~\ref{Esurf}(b)], whereas the other branch is for a low density oblate state [i.e.~$w_z\ll w_\rho$, see right inset to Fig.~\ref{Esurf}(a)]. For clarity we will refer to these as the high-density phase (HDP) and the low-density phase (LDP) respectively.

The form of  the LDP solution is dominated by the interplay of two-body interactions and the trap potential (cf.~the Thomas Fermi limit \cite{Dalfovo1999}), and thus corresponds to the typical regime of Bose-Einstein condensates realised in experiments. Notably, in this regime if the radial trap confinement $\omega_\rho$ is allowed to go to zero then this solution will broaden ($w_\rho\to\infty$) and approach a uniform state of zero density.

The HDP occurs when the DDI is strong compared to the s-wave interactions, i.e.~necessarily in the dipole dominated regime defined by $a_{\mathrm{dd}}>a_s$.  The HDP is a dense droplet arising from the competition between the DDIs (that tends to collapse the condensate to a high density spike that is elongated along $z$), the repulsive TBI (that acts to oppose the droplet density getting too high) and the $z$ confinement (that opposes the droplet extending too far along $z$). Unlike the LDP, the HDP state remains essentially unchanged as the radial trap frequency is reduced, although we emphasise that the $z$ confinement must remain. Thus this state is in effect a quasi-two-dimensional bright soliton. It differs from the two-dimensional bright soliton predicted by Pedri \textit{et al.}~\cite{Pedri2005a}, which  requires a negative DDI and tight (quasi-two-dimensional) trapping.

An interesting feature of the LDP and HDP branches observed in Fig.~\ref{Esurf}(a) is that they intersect at a \textit{transition point} ($a_s\approx90\,a_0$, where $a_0$ is the Bohr radius). For values of $a_s$ less that this the HDP state is the global energy minimum (i.e.~ground state), whereas for greater values of $a_s$  the LDP state is the global energy minimum. The LDP and HDP branches both extend into $a_s$ regions where they are not the ground state, and here they are meta-stable states. For values of $a_s$ sufficiently far from the transition point the meta-stable states end [i.e.~the local minimum eventually vanishes, e.g.~see Fig.~\ref{Esurf}(c) and (e)]. This general behavior is that of a first order phase transition, and we discuss this further in Sec.~\ref{secPD}.

\begin{figure}[htbp]
 \vspace*{-0.5cm}
   \centering
 \includegraphics[width=3.4in]{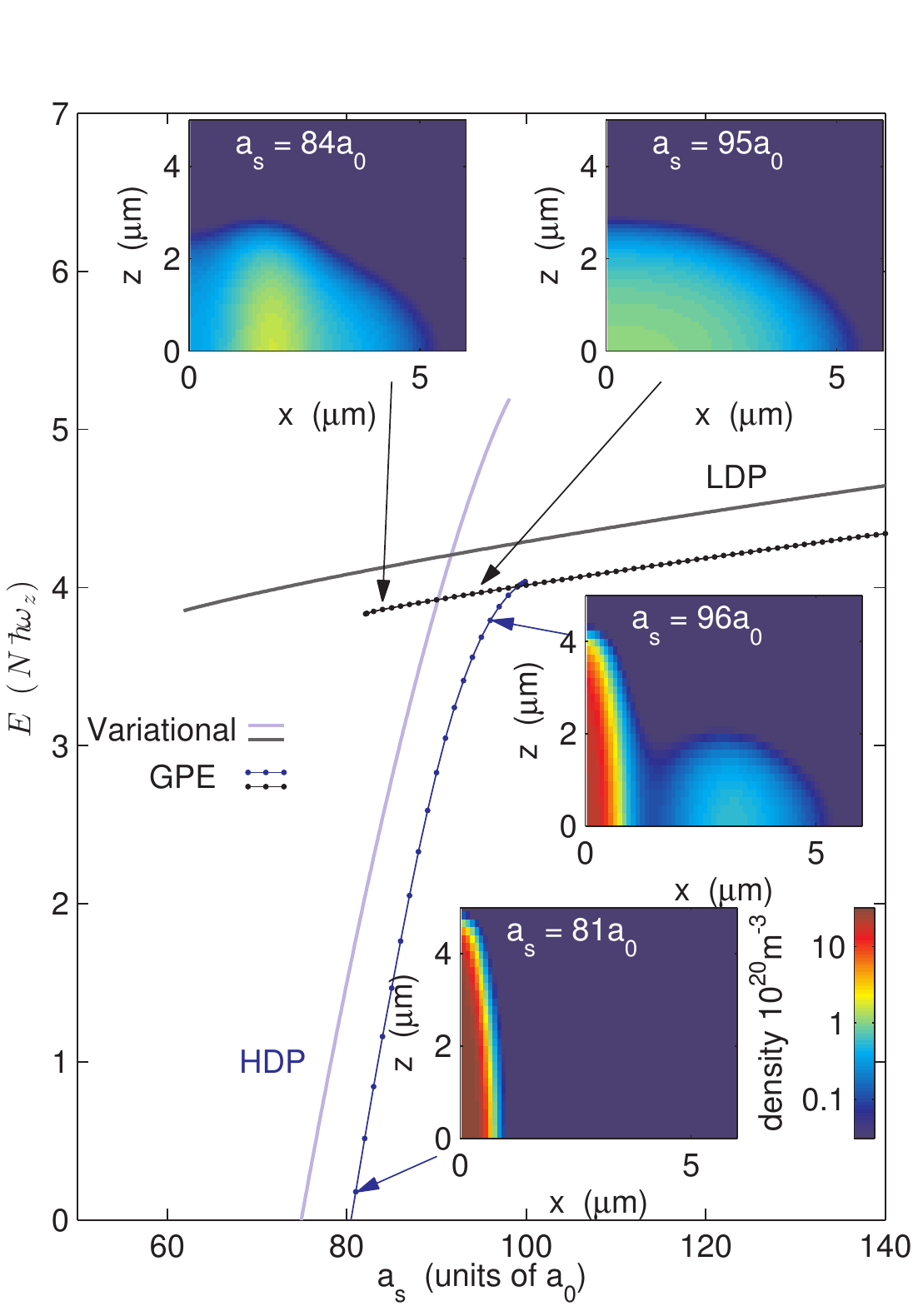} 
   \vspace*{-1cm}\caption{(Color online)   GPE solutions (dotted lines) compared to the variational solutions (light solid lines). Density slices $|\psi(x,0,z)|^2$ of the GPE solutions are shown in the insets for several parameter values. Other parameters as in Fig.~\ref{Esurf}.}
   \label{GPEsolns}
\end{figure}

It is useful to investigate the predictions of the variational solutions by comparing them to full numerical solutions of the GPE (\ref{tiGPE}). To do this we find a solution on the LDP branch at a large initial value of $a_s$   (i.e., $a_s=140\,a_0$) using the variational solution as an initial guess for our GPE solver. We then follow the branch by  decreasing the value of $a_s$  by an amount $\Delta a_s$ and  using the previous solution as an initial guess for the GPE solver. Eventually we can no longer find a solution, despite decreasing the size of the $a_s$ steps to $|\Delta a_s|\sim10^{-3}a_0$, which we interpret as the end of the branch. When this happens a quasi-particle excitation softens (approaches zero energy), which marks the onset of a dynamical instability. Similarly, starting from a low value of $a_s$ we obtain a solution on the lower (HDP) branch, and can follow this up by slowly increasing $a_s$. The energy of the two branches we obtain from this procedure, and some example states, are shown in Fig.~\ref{GPEsolns}. These results show that the variational solution accurately predicts the qualitative behaviour, although tends to underestimate the transition point, i.e.~the $a_s$ value where the two branches cross.

A few features of the full GPE wavefunctions (see insets to Fig.~\ref{GPEsolns}) are worth noting: i) the LDP solution for $a_s=84\,a_0$, near the end of the LDP branch, has a local density minimum at $\rho=0$. Such ``density oscillating" ground states are known to occur in dipolar condensates for specific trap geometries and interaction parameter regimes (e.g.~see \cite{Ronen2007a,Lu2010a,Bisset2012,Martin2012a}). ii) the HDP solution for $a_s=96\,a_0$, near the end of the HDP branch, has a halo-like ring in the radial plane. This occurs because there is a shallow minimum in the effective potential at a finite radius (outside the main condensate) in the $xy$-plane that can allow the condensate to tunnel into it (see discussion of the ``Saturn-ring  instability" in Ref.~\cite{Eberlein2005a}).

\subsection{Phase diagram and dynamics}\label{secPD}
\begin{figure}[htbp]
   \centering
 \includegraphics[width=3.3in]{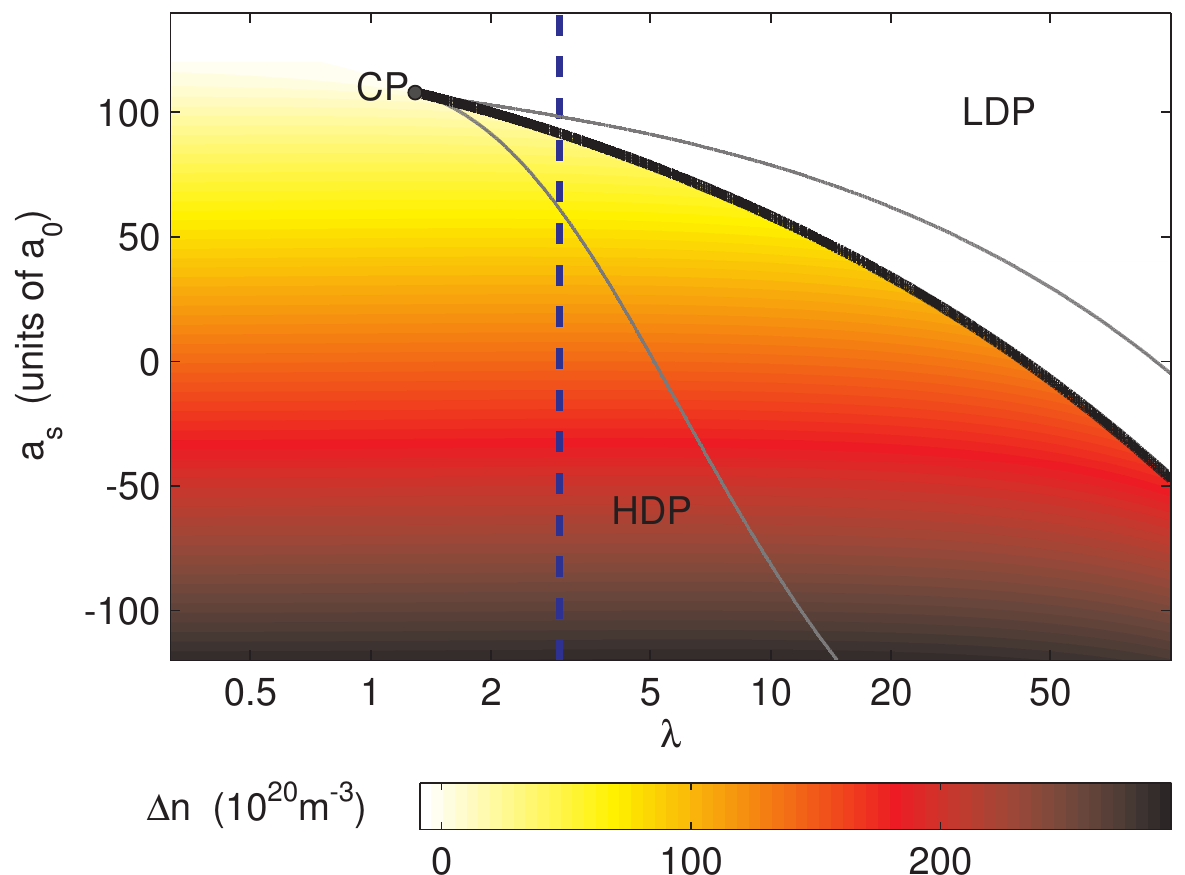} 
   \caption{(Color online)  Phase diagram as a function of s-wave scattering length and trap aspect ratio $\lambda=\omega_z/\omega_\rho$ for a system of fixed geometric mean trap frequency  [$\bar\omega=(\omega_\rho^2\omega_z)^{1/3}$]. The colors indicate the peak density difference between the ground state solution at that point in the phase diagram and the critical point solution (i.e.~$\Delta n$). The solid black line marks the transition point where the LDP and HDP solutions are energy degenerate, and the thin grey lines indicate the spinodal lines where the metastable states terminate.   CP marks the critical point where the transition line terminates. The vertical dashed line indicates the parameter regime considered in Fig.~\ref{Esurf}. Results are determined using the gaussian variational approximation. Other parameters as in Fig.~\ref{Esurf}.}
   \label{varPD}
\end{figure}
The LDP and HDP states can be characterized by their peak density which (for the variational ansatz) always occurs at trap centre. 
We can then produce a phase diagram for this system using the density difference
\begin{equation}
\Delta n=n_\mathrm{peak} -n_\mathrm{peak}^\mathrm{CP},
\end{equation} 
as the order parameter, following a standard convention for the liquid-gas phase transition (where density also serves as the distinguishing characteristic of the two phases). Here $n_\mathrm{peak}$ is the peak density of the ground state (for the parameters under consideration), while $n_\mathrm{peak}^\mathrm{CP}$ is the peak density at the critical point (which we identify below).  We show the results of this analysis in Fig.~\ref{varPD} as function of the s-wave scattering length and the trap aspect ratio. The shading indicates the value of the order parameter and we mark a transition line on the phase diagram where the energies of the two minima are degenerate [i.e.~where the two branches intercept, as seen in Fig.~\ref{Esurf}(a)]. This transition line coincides with a jump in $\Delta n$. This transition line terminates at a critical point at a nearly isotropic trap, where the density difference  goes to zero. 
 
\begin{figure}[htbp]
   \centering
 \includegraphics[width=3.4in]{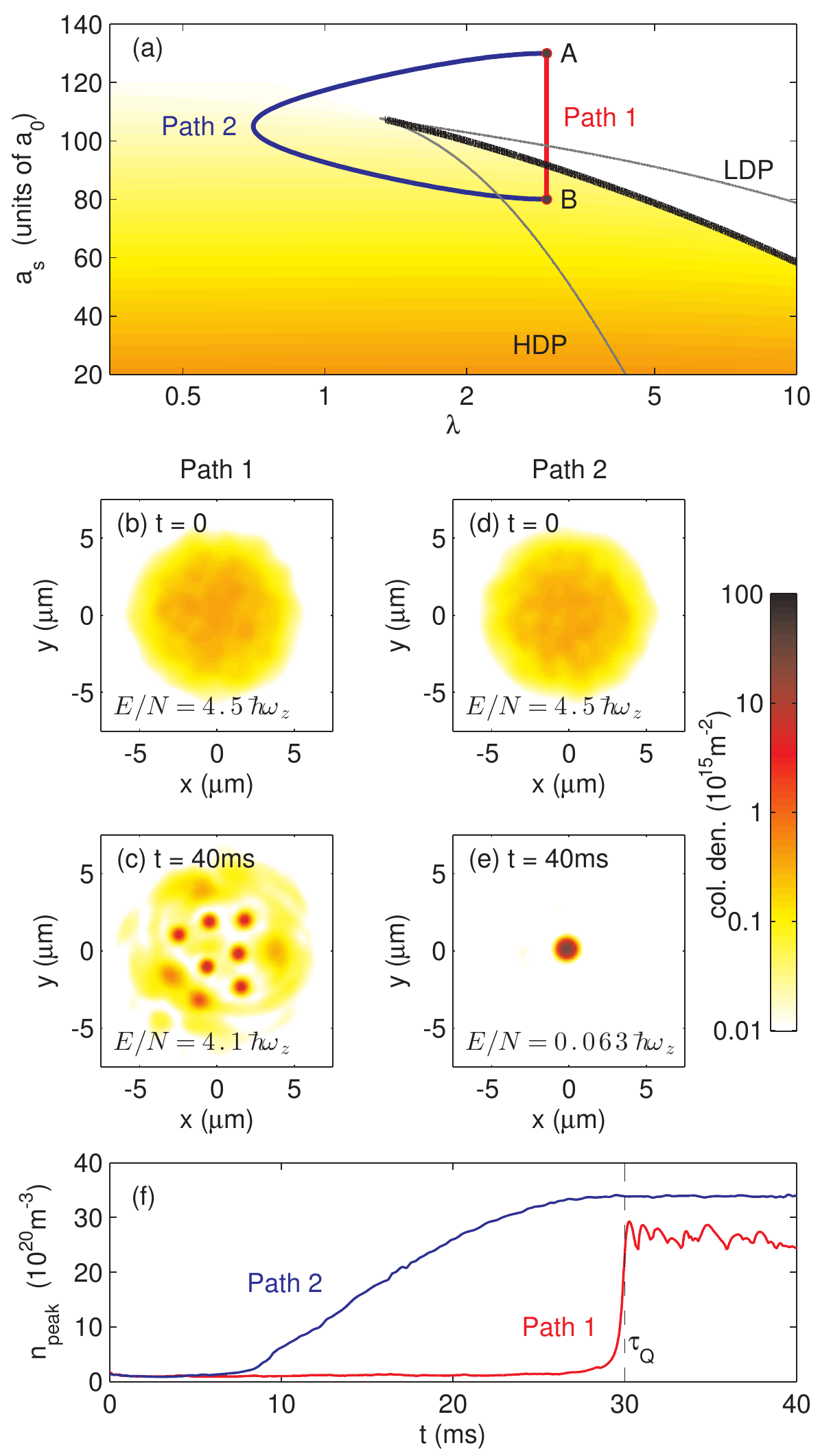} 
   \caption{(Color online)  Pathways to the HDL phase and formation dynamics. (a) The phase diagram from Fig.~\ref{varPD} with pathways added to show the two processes (path 1 and path 2) we use to take the system from phase diagram location A (coordinates: $a_s=130\,a_0$, $\lambda=2.96$) to location B (coordinates: $a_s=80\,a_0$, $\lambda=2.96$).  The paths are traversed over a  time of   $\tau_Q=30\,$ms.  Simulation along path 1: (b) Initial state  and  (c) final state 10$\,$ms after the process is complete (i.e.~at $t=\tau_Q+10\,$ms).  Simulation along path 2:  (d) Initial state  and  (e) final state  10$\,$ms after the process is complete. (f) The evolution of the peak density of the system during the evolution. Three-dimensional density isosurfaces of the states in (c) and (e) are shown in Fig.~\ref{densurf}(a) and (b), respectively. Other parameters as in Fig.~\ref{Esurf}.}
   \label{paths}
\end{figure}

We now consider the dynamics of the system from a location in the LDP region of the phase diagram [labelled A] to a location in the HDP region [labelled B] as indicated in Fig.~\ref{paths}(a). These locations have the same trap potential, but differ in the value of the s-wave scattering length. We effect a process to take us between locations A and B by changing the relevant system parameters, namely the s-wave scattering length (that can be adjusted using Feshbach resonances) and the trap parameters (that can be adjusted through control of the externally applied light fields).  We simulate the system dynamics, including quantum and thermal effects, using the time-dependent GPE (\ref{tdGPE}) with noisy initial conditions chosen according to the truncated Wigner prescription. Details of the Wigner method and other aspects of the simulation are discussed in the Appendix.
 
 We consider two distinct process paths to bring the system from A to B (precise details of these paths are provided in the Appendix).
 The first path [path 1 in Fig.~\ref{paths}(a)] corresponds to a linear quench of the s-wave scattering length from the initial value of $a_i=130\,a_0$ to the final value of $a_f=70\,a_0$ over a duration of $\tau_Q=30\,$ms (while holding the trap constant). This time scale for performing the process is chosen to be longer than the trap period ($2\pi/\bar{\omega}=15.5\,$ms).
 The simulation results [see Fig.~\ref{paths}(b), (c)] reveal that the system forms a crystal of droplets, rather than the ground state configuration of a single droplet at trap centre. This occurs because this path crosses the first order phase transition at a finite rate.  The system remains in the metastable LDP state (hence with excess energy) until about $t\approx29\,$ms when the droplets locally nucleate. This is revealed by examining the dynamics of the peak density [see Fig.~\ref{paths}(f)], which suddenly increases when the droplets form.
 The second path [path 2 in Fig.~\ref{paths}(a)] is an elliptical path that goes around the critical point, thus avoiding the need to cross the first order phase transition line. This path is also traversed over a duration of  $\tau_Q=30\,$ms. In this case the simulation [see Fig.~\ref{paths}(d), (e)] reveals that the system forms a single droplet at trap centre, and is close to the expected HDP ground state. The final energy of the simulation on this path is much lower that that for path 1 [energies given in Figs.~\ref{paths}(b)-(e)], so that we verify there is significantly less heating along this path. We also see from the evolution of the peak density [Fig.~\ref{paths}(f)], that the single droplet forms quite smoothly as the path is traversed. 
 
 We have simulated  other paths like those discussed above and have investigated the effect of time scale $\tau_Q$, trap geometry, and the final value of $a_s$ on the dynamics. We find that for $\tau_Q=10\,$ms a single droplet can form on path 2, however more energy (heating) is added through the excitement of collective excitations via the more rapid change in trap geometry  and interaction strength. The linear quench we study (path 1) is similar to that used to reproduce the experimental observations made in Ref.~\cite{Kadau2016a} (the main difference is that there  $\tau_Q=0.5\,$ms  was used). Qualitatively similar dynamics to that of  Fig.~\ref{paths}(b) and (c) is found for path 1 with $\tau_Q=0.5\,$ms, except that there is more heating. 
 We have checked to see if crossing the phase transition line much more slowly could lead to the formation of a single droplet. Performing simulations along path 1 but using $\tau_Q=100\,$ms we still observe a crystal to form. For a shallower quench to $a_s=90\,a_0$ (i.e.~just over the transition point for $\lambda=2.96$, see Fig.~\ref{GPEsolns}) we find that the LDP state is metastable, and it can take  $\sim10\,$ms for the crystalization to occur (post quench), with the precise time depending on the quench rate and temperature of the initial state. For deeper quenches the crystal tends to nuclear much faster, and more droplets form.

\section{Conclusion} 
In this paper we have developed a phase diagram for dipolar condensate with TBIs. This work provides a global view of the  specific dynamics for this system presented in references \cite{Xi2016a,Bisset2015a}.
Our results make it clear that the crystallization process observed in those studies was the result of a crossing a first order phase transition nonadiabatically. Importantly, we show that by going around the critical point it is possible to follow the ground state adiabatically, and thus produce a single droplet. While our results have focused on a particular parameter regime motivated by recent experiments, we demonstrate that the variational treatment provides a reasonably accurate model that could be easily deployed to   other regimes.

A number of directions present themselves for future work. First, the treatment we present here could be adapted to the quantum fluctuation mechanism that has been proposed as an alternative explanation for stabilising the crystalline phase \cite{Ferrier-Barbut2016a,Wachtler2016a}. In these studies a term that contributes  to the energy density with  $n^{5/2}$ scaling is introduced, motivated by results of Lima \textit{et al.}~\cite{Lima2011a,Lima2012a} (cf.~$n^3$ for the TBI).

Another direction will be to extend the theory towards flatter (quasi-two-dimensional) systems where rotons are expected to play a more obvious role near the point of instability for the LDP (e.g.~see \cite{Santos2003a,Ronen2006a,*Blakie2012a,*Corson2013a,*Boudjemaa2013a,*JonaLasinio2013,*Bisset2013a,Baillie2015a}). 

 \section{Acknowledgments}
 We gratefully acknowledge valuable discussions with R.~Bisset and R.~Wilson,  the contribution of NZ eScience Infrastructure (NeSI) high-performance computing facilities, and support from the Marsden Fund of the Royal Society of New Zealand.   
   
\appendix
 \section*{Appendix: Simulation details}
The  procedure that we use for conducting the dynamical simulations reported in Sec.~\ref{secPD}  is similar to that used in Ref.~\cite{Bisset2015a}. The initial state is based on the  solution  $\psi_0(\br)$ to  Eq.~(\ref{tiGPE}) for $N=15,000$ atoms with $a_s=130\,a_0$, and other parameters as Fig.~\ref{Esurf}. This state is obtained by using a Newton-Krylov scheme (see \cite{Ronen2006a,Martin2012a}).  To mimic the effects of quantum and thermal fluctuations we add initial state fluctuations, which play an important role in seeding the droplet formation dynamics. To be precise, these are added as 
\begin{equation}
\psi(\br,0)=\psi_0(\br)+{\sum_{n}}'\alpha_n\phi_n(\br),\label{noise}
\end{equation}
where $\epsilon_n\phi_n=H_{\mathrm{sp}}\phi_n$ are the single particle eigenstates (harmonic oscillator basis), the coefficients $\alpha_n$ are complex gaussian random variables with  
\begin{equation}
\langle |\alpha_n|^2\rangle=\frac{1}{e^{\epsilon_n/k_BT}-1}+\frac{1}{2}.\label{alphan}
\end{equation}
The notation $\sum'$ denotes that the summation in (\ref{noise}) is restricted to modes with energy $\epsilon_n\le 2k_BT$. This choice of fluctuations is consistent with the truncated Wigner prescription (see \cite{Steel1998a,cfieldRev2008}) for a system at temperature $T$. The  results we present in Fig.~\ref{paths} are for $T= 20\,n$K, adding approximately 400 atoms to the system (cf.~the ideal condensation temperature of $T_c=72\,n$K for $N=15\times10^3$).  We also note that the $\frac{1}{2}$ term in Eq.~(\ref{alphan}) accounts for quantum fluctuations in the initial state. As we add this in the single particle basis, rather than the Bogoliubov quasiparticle basis, it is not a comprehensive treatment of the quantum fluctuations in the system.

 For dynamics we evolve the system according to the GPE (\ref{tdGPE}) discretised on a three-dimensional grid in a cubic box of dimension $23.4\,\mu$m, with grid point spacing of $\Delta x=0.1672\,\mu$m (i.e.~140 points in each direction). The time-dependent GPE is propagated in time using a 4th order Runge-Kutta method.  The s-wave interaction and trap parameters are changed during the time interval $0<t<\tau_Q$ (and thereafter  held constant) in the evolution according to the path taken. 
For path 1:
 \begin{align}  
 a_s(t)&=    
       a_i+(a_f-a_i)\frac{t}{\tau_Q},  \\
 \lambda(t)&=\lambda_i.
 \end{align}
For path 2:
  \begin{align}  
 a_s(t)&=    \frac{a_i+a_f}{2}+\frac{a_i-a_f}{2}\cos\left(\frac{\pi t}{\tau_Q}\right),\\ 
 \lambda(t)&=\lambda_i+\lambda_r\sin\left(\frac{\pi t}{\tau_Q}\right).
 \end{align}
 For both paths the initial and final scattering lengths are $a_i=130\,a_0$ and $a_f=80\,a_0$, and the initial (and final) trap aspect ratio is $\lambda_i=2.96$. For path 2
we also use $\lambda_r=2.25$, and we note that because the phase diagram is for fixed geometric mean trap frequency $\bar\omega$, the trap frequencies are adjusted according to $\omega_\rho(t)=\bar\omega/\lambda(t)^{1/3}$ and $\omega_z(t)=\bar\omega\lambda(t)^{2/3}$.

\end{document}